\documentclass[journal]{IEEEtran}

\usepackage{amsmath}

\ifCLASSINFOpdf
  \usepackage[pdftex]{graphicx}
\else
  \usepackage[dvips]{graphicx}
\fi

\usepackage{array}
\usepackage[printonlyused]{acronym}
\usepackage{cite}
\usepackage[tight]{subfigure}
\usepackage{multirow}
\usepackage{xcolor}
\usepackage{soulpos}
\ulposdef{\hlc}[xoffset=1pt]{\mbox{\color{cyan!30}\rule[-.8ex]{\ulwidth}{3ex}}}
\ulposdef{\hlg}[xoffset=1pt]{\mbox{\color{green!30}\rule[-.8ex]{\ulwidth}{3ex}}}

\usepackage{color, soul}
\usepackage{upgreek}

\renewcommand{\hl}[1]{#1}
\renewcommand{\hlg}[1]{#1}

\title{Graphene-based Wireless Agile Interconnects for Massive Heterogeneous Multi-chip Processors}

\author{Sergi Abadal, Robert Guirado, Hamidreza Taghvaee, Akshay Jain, Elana Pereira de Santana,\\Peter Haring Bol\'{i}var, Mohamed Saeed, Renato Negra, Zhenxing Wang, Kun-Ta Wang, Max C. Lemme,\\Joshua Klein, Marina Zapater, Alexandre Levisse, David Atienza, Davide Rossi, Francesco Conti,\\Martino Dazzi, Geethan Karunaratne, Irem Boybat and Abu Sebastian
\thanks{Sergi Abadal (abadal@ac.upc.edu), Robert Guirado, Hamidreza Taghvaee, and Akshay Jain are with the Universitat Polit\`{e}cnica de Catalunya. Elana Pereira de Santana and Peter Haring Bol\'{i}var are with the University of Siegen. Mohamed Saeed, Renato Negra, Kun-Ta Wang, and Max C. Lemme are with the RWTH Aachen University. Zhenxing Wang, Kun-Ta Wang, and Max C. Lemme are also with AMO GmbH. Joshua Klein, Marina Zapater, Alexandre Levisse, and David Atienza are with the Swiss Federal Institute of Technology Lausanne. Marina Zapater is also with University of Applied Sciences and Arts Western Switzerland. Davide Rossi and Francesco Conti are with the University of Bologna. Martino Dazzi, Geethan Karunaratne, Irem Boybat and Abu Sebastian are with IBM Research Europe.}
}

\acrodef{RC}{Resistive-Capacitive}
\acrodef{CMOS}{Complementary Metal--Oxide--Semiconductor}
\acrodef{mm-Wave}{millimeter-Wave}
\acrodef{WNSN}{Wireless NanoSensor Network}
\acrodef{WSN}{Wireless Sensor Network}
\acrodef{MAC}{Medium Access Control}
\acrodef{QoS}{Quality of Service}
\acrodef{TS-OOK}{Time Spread On-Off Keying}
\acrodef{CSMA}{Carrier Sense Multiple Access}
\acrodef{GWNoC}{Graphene Wireless Network-on-Chip}
\acrodef{WNoC}{Wireless Network-on-Chip}
\acrodef{NoC}{Network-on-Chip}
\acrodef{CMP}{Chip Multiprocessor}
\acrodef{TDMA}{Time Division Multiple Access}
\acrodef{FDMA}{Frequency Division Multiple Access}
\acrodef{CDMA}{Code Division Multiple Access}
\acrodef{ACK}{Acknowledgment message}
\acrodef{RF}{Radio-Frequency}
\acrodef{IR}{Impulse Radio}
\acrodef{OOK}{On-Off-Keying}
\acrodef{BER}{Bit Error Rate}
\acrodef{DVFS}{Dynamic Voltage and Frequency Scaling}
\acrodef{ARQ}{Automatic Repeat reQuest}
\acrodef{FEC}{Forward Error Correction}
\acrodef{M2M}{Machine-to-Machine}
\acrodef{mmWave}{millimeter-wave}
\acrodef{THz}{Terahertz}
\acrodef{$f_{max}$}{maximum oscillation frequency}
\acrodef{$f_T$}{cutoff frequency}
\acrodef{FET}{field effect transistor}

\begin{document}

\maketitle

\begin{abstract} 
The main design principles in computer architecture have recently shifted from a monolithic scaling-driven approach to the development of heterogeneous architectures that tightly co-integrate multiple specialized processor and memory chiplets. In such data-hungry multi-chip architectures, current Networks-in-Package (NiPs) may not be enough to cater to their heterogeneous and fast-changing communication demands. This position paper makes the case for wireless in-package networking as the enabler of efficient and versatile wired-wireless interconnect fabrics for massive heterogeneous processors. To that end, the use of graphene-based antennas and transceivers with unique frequency-beam reconfigurability in the terahertz band is proposed. The feasibility of such a wireless vision and the main research challenges towards its realization are analyzed from the technological, communications, and computer architecture perspectives. 
\end{abstract}


\acresetall

\section*{\Large \textbf{Introduction}} 

The end of Dennard scaling has led to the rise of multicore processors, whereby a set of independent \emph{cores} are integrated within a single chip. In theory, the potential of these processors scales with the number of cores \cite{Bohnenstiehl2017}. In practice, realizing such a potential requires an on-chip interconnect capable of providing high-throughput and low-latency data sharing among cores. This fundamentally implies that communication, and not computation, becomes the main bottleneck in these multicore chips, especially as we scale towards thousand cores \cite{Nychis2012}.

\begin{figure}
  \centering
  \includegraphics[width=\columnwidth]{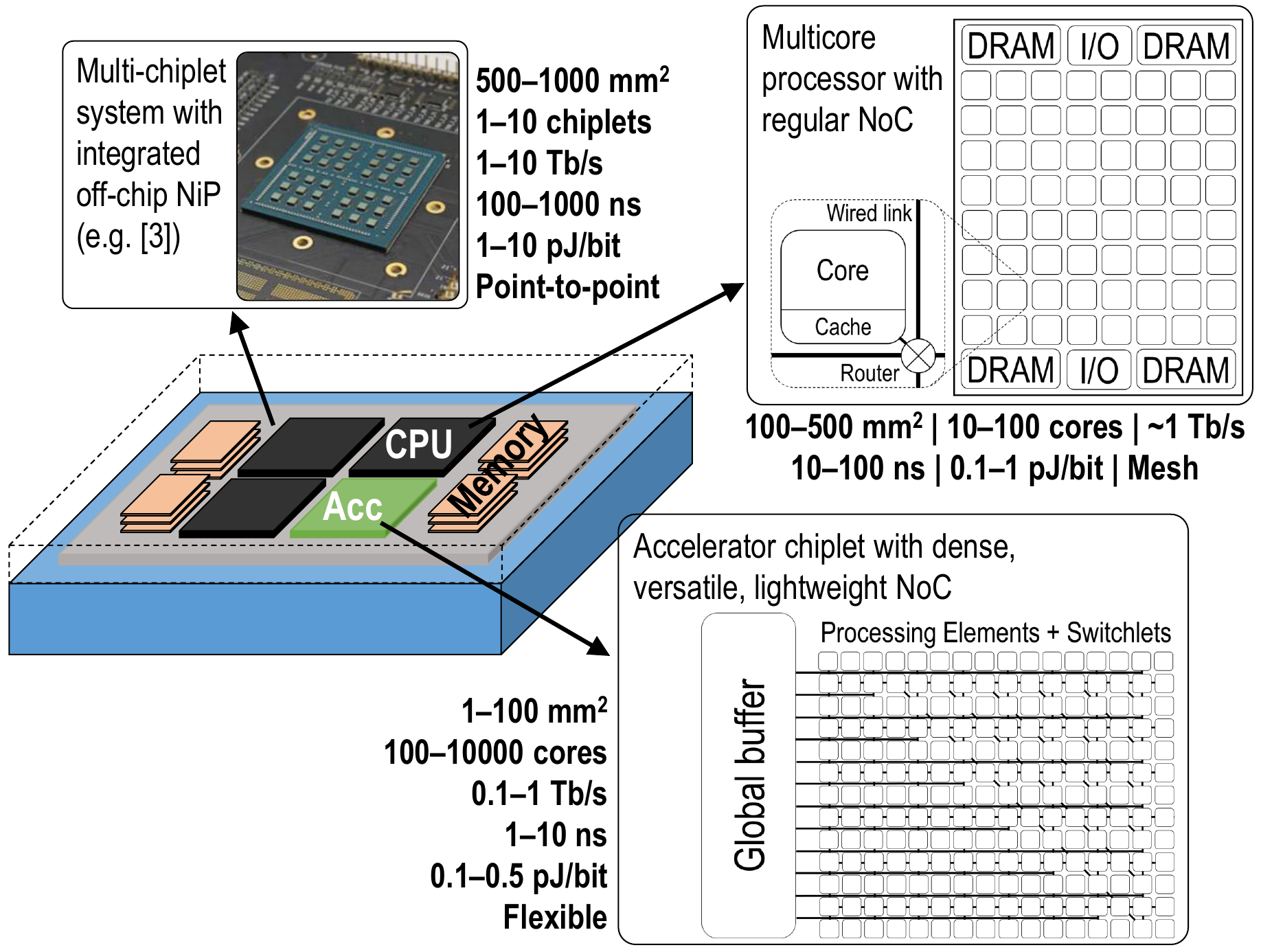}
	\caption{The chip-scale communication landscape in the heterogeneous chiplet era: Network-in-Package (NiP) to interconnect chiplets, Network-on-Chip (NoC) for multicore processors, and dense fabrics for accelerators. For the three scenarios, we list popular system sizes, number of nodes, bisection bandwidth, latency, energy per transmitted bit, and topology.}
  \label{fig:intro}
\end{figure}

Current architectural trends maintain the multicore essence, but have recently resorted to disintegration and specialization to continue improving performance. Disintegration means that large chips are being replaced by ensembles of smaller, heterogeneous chiplets that are interconnected through a silicon interposer within a System-in-Package (SiP), as illustrated in Fig. \ref{fig:intro}. Specialization means that chiplets are becoming increasingly optimized to accelerate a given particular computation. This has two key implications in the communications requirements. 
At the system level, communication becomes dominated by off-chip transfers and mandated by the heterogenous mixture of chiplets of a particular SiP. At the chiplet level, some accelerators are extremely data-intensive and require an ultra-dense on-chip interconnect, further extending the spectrum of communication needs of SiPs. Artificial Intelligence (AI) accelerators are a very popular example of this trend \cite{Shao2019}.

To satisfy the communication needs of multicore chips, Network-on-Chip (NoC) has been the standard interconnect fabric in the last decade \cite{Nychis2012}. NoCs are packet-switched networks of integrated routers and wires typically arranged in a mesh topology. This approach, however, has important limitations when scaled beyond tens of cores. In particular, chip-wide and broadcast transfers suffer from increasing latency and energy consumption due to the many hops needed to reach the destination(s) \cite{Nychis2012}. \hl{This problem cannot be fixed with longer interconnects due to the dramatic growth in resistivity of copper wires in deeply scaled technology nodes} \cite{Todri-Sanial2017}. Thus, when developing ultra-dense accelerators, the NoC is forced to become a specialized fabric with \hl{short wires and} little flexibility. Further, when moving to multi-chip SiPs, the concept of NoC has been extended to incorporate off-chip links and form a Network-in-Package (NiP) \cite{Shao2019}, aggravating their existing issues and adding new ones. For instance, NiPs are rigid and prone to over-provisioning in a scenario that demands a versatile solution to address heterogeneous communication needs. This, together with the connectivity constraints of each chiplet determined by the amount of physical connection pins, severely limits the applicability and scalability of the SiP approach. 
 
\hl{Although carbon-based materials have been proposed to alleviate some of the limitations of electrical interconnects} \cite{Todri-Sanial2017}\hl{, emerging technologies such as nanophotonics or wireless interconnects have been proposed in the last decade to address the issues of NoCs} \cite{Karkar2016}. Now, given the advent of accelerators and the increasing importance of off-chip links in chiplet-based architectures, there is a renewed interest in these technologies. In particular, nanophotonic networks have been proposed due to their massive bandwidth density and high energy efficiency \cite{Wade2020a}. While this approach solves the bandwidth problem of NiPs, \hl{nanophotonic technology suffers from high cost and design complexity due to the current unavailability of affordable integrated light sources and the need for conversions between the electrical and optical domains}. Moreover, its flexibility is still limited by the need for laying down waveguides across the system.
Instead, the wireless approach enabled by recent advances in integrated millimeter-wave (mmWave, 30--300 GHz) devices is theoretically capable of providing the desired reconfigurability and, thus, \hl{becomes a natural complement} to the rigid yet effective wired networks \cite{Shamim2017}.

In this paper, we make the case for wireless communications within a multi-chip package as a key element for efficient and scalable NiP-based architectures. While a first implementation of the wireless approach could be done with conventional \ac{CMOS} mmWave technology, the use of graphene-based antennas and transceivers in the terahertz band (0.3--3 THz) is seen as an excellent opportunity in this context due to various reasons summarized in Table \ref{tab:reqs}. At the antenna, graphene offers unique tunability and miniaturization properties, which enable the realization of compact tunable beam-steering arrays \cite{singh2020design}. At the transceiver, graphene may enable ultra-high modulation speeds at low cost by virtue of its high responsivity, excellent linearity, and wide dynamic range \cite{Saeed2021}. 
With an appropriate protocol stack, graphene's tunability can be translated into architectural reconfigurability, which is a very appealing aspect in multi-chip systems \cite{Shao2019, Shamim2017}, where the ample bandwidth in the THz range could help satisfy the needs of massive architectures with data-intensive accelerators (see Fig. \ref{fig:intro}).

\begin{table*}[!t] 
\caption{Comparison of different interconnect technologies for Network-in-Package (NiP).\break Blue boxes indicate the best alternative in each particular metric.\label{tab:reqs}}
\vspace{-0.2cm}
\centering
\begin{tabular}{c|ccc|c} 
\hline
{\bf Metric} & {\bf Interposer} & {\bf Optics} & {\bf Wireless} & {\bf Graphene} \\
\hline
Medium & Wires & Waveguides & Package & Package \\
Frequency & Baseband & Optical & Millimeter Wave & Terahertz \\
Latency & 10--100 ns & 10--100 ns & \hlc{1--10 ns} & \hlc{1--10 ns} \\
Bisection Bandwidth & 0.1--1 Tb/s & \hlc{1--100 Tb/s} & 0.01--0.1 Tb/s & 0.1--1 Tb/s\\
Energy Efficency & 1--10 pJ/bit & \hlc{0.1--10 pJ/bit} & 1--10 pJ/bit & 1--10 pJ/bit \\
\hl{Scaling Mechanism} & \hlc{More wires} & \hlc{More waveguides} & Wider spectrum & Frequency-space channels \\
Multiplexing & No & Time* & Time & \hlc{Space, time, frequency} \\
\hl{Architectural Impact} & Low  & Low   & Medium  & \hlc{High}   \\
\hl{Broadcast Capability} & Poor & Expensive & \hlc{Native} & \hlc{Native} \\
\hl{Implementation Cost} & \hlc{Moderate} & High  & Medium   & Medium   \\
\hl{CMOS Compatibility} & \hlc{High/BEOL$^{\dagger}$} & Low/BEOL$^{\dagger}$   & High/FEOL$^{\diamond}$  & High/FEOL$^{\diamond}$ \\
\hl{Design Complexity} & \hlc{Low}   & High    & Medium    & \hlc{Low}    \\
\hline
\end{tabular}
\\ *only if global waveguides are used. $^{\dagger}$Back end of line. $^{\diamond}$Front end of line.
\vspace{-0.3cm}
\end{table*}

This work aims to provide a description of the envisaged networking scenario (Fig. \ref{fig:vision}) from the technological, communications, and architectural perspectives. We first briefly discuss relevant literature in integrated graphene antennas and transceivers. We also depict the structure of a typical multi-chip package and its wireless propagation characteristics, to then outline the roles of the wireless communications plane within the architecture. Finally, we describe the open issues and research challenges towards the realization of this in-package networking vision, from the technological integration aspect up to the heterogeneous architecture design and simulation. \hl{Hence, in summary, the main contributions of this article are:}
\begin{itemize}
    \item \hl{The vision of agile wireless networks for multi-chip architectures is laid down, providing a description of the state of the art of its main enabler: graphene RF technology.}
    \item \hl{The wireless chip-scale communications context is analyzed, detailing the main physical constraints, communication requirements, and architectural impact of the proposed vision.}
    \item \hl{The path towards the realization of the wireless agile chip-scale networks is described by outlining the technology, implementation, communications, and architecture challenges to be overcome.}
\end{itemize}

\section*{\Large \textbf{Agile Wireless Interconnect Fabrics enabled by Graphene}} 

\begin{figure}
   \centering
   \includegraphics[width=\columnwidth]{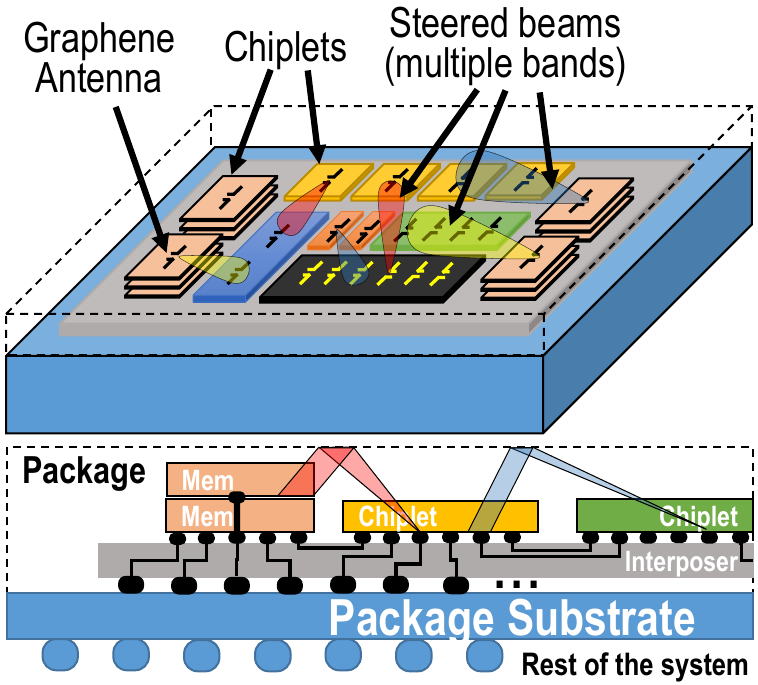}
 	\caption{Schematic diagram of a System-in-Package (SiP) hosting a heterogeneous set of chiplets. The interconnect fabric is composed of a silicon interposer Network-in-Package (NiP) augmented with graphene-based agile wireless links.}
   \label{fig:vision}
 \end{figure}

Fig. \ref{fig:vision} illustrates the proposed wireless networking vision within the context of heterogeneous multi-chip architectures. In this paradigm, graphene-based antennas and transceivers are co-integrated with the computing elements of a SiP and use the system package as the wireless propagation medium. The wireless links form a network within and across chiplets that complements the wired interconnect fabric.

In this context, wireless communications offer multiple benefits. By not needing to lay down wires between source and destination, wireless technology is able to bypass I/O pin limitations or wire routing constraints. The newly available bandwidth can be shared dynamically to adapt to the needs of the architecture. The latency, critical in this scenario, can be reduced by an order of magnitude, especially in transfers that would otherwise require the traversal of long multihop paths within the dense maze of wired interconnects. A particularly relevant example is broadcast, which implies flooding the network in wired NiPs. In the wireless case, broadcast becomes scalable since additional chiplets only need to incorporate a transceiver to participate in the communication.  

The use of graphene antennas turns the wireless interconnect into a wireless network. Such compact beam-steerable, frequency-tunable antennas lead to substantial improvements in antenna gain, achievable capacity, and network-level flexibility. These additional benefits can be exploited by the architecture through a judicious orchestration of the space, time, and frequency wireless channels. We next describe \hl{recent advances enabling our vision, together with its} most salient characteristics at the antenna, RF circuit, network, and architecture fronts.

\subsection*{\centering \scshape \textbf{Graphene-based Antennas \hl{and RF Circuits}}}

The concept of wireless flexibility in this scenario can only be realized by means of a compact, high-bandwidth, highly-reconfigurable wireless technology. Graphene is ideally suited to this purpose thanks to its outstanding optoelectronic properties. Through the support of Surface Plasmon Polaritons (SPPs) in the THz band, graphene opens the door to the design of THz antennas with ultra-broad tunable resonance and reconfigurable beams. Although full-fledged prototypes are still not available, the potential of graphene antennas has been proven in multiple works \cite{wang2019graphene, singh2020design}. Moreover, the integration of graphene to mature process platforms such as \ac{CMOS} has been demonstrated with great success \cite{neumaier2019integrating}. Therefore, graphene antennas can be integrated into conventional processor chips.

\begin{figure*}
   \centering
   \includegraphics[width=\textwidth]{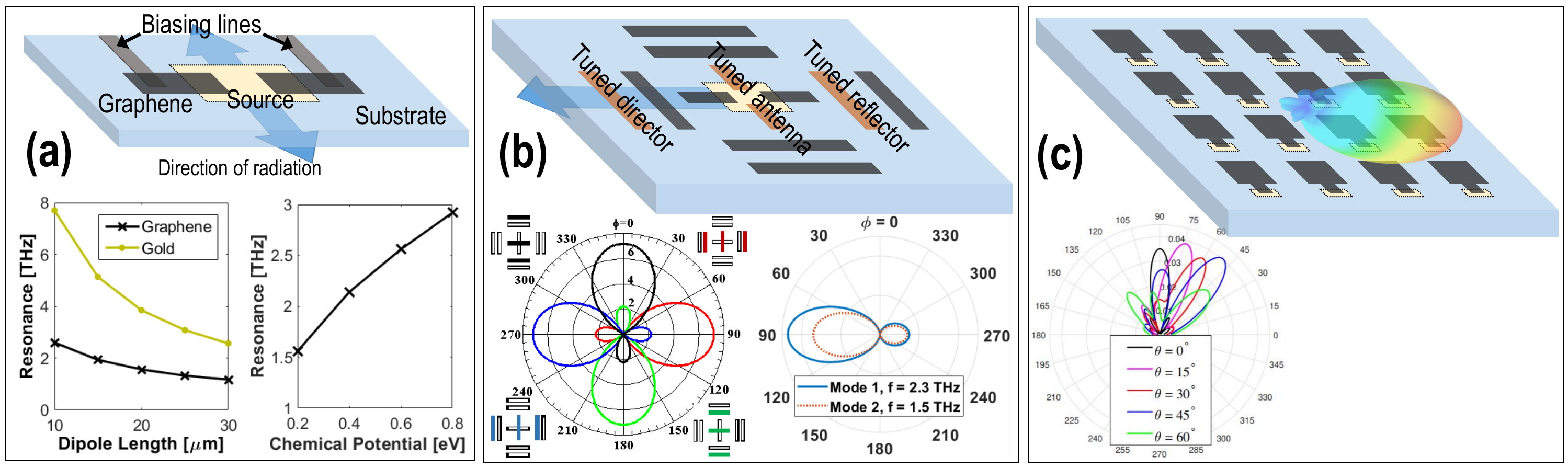}
 	\caption{Evolution of graphene antennas from (a) simple graphene dipoles illustrating miniaturization and tunability, to (b) compact Yagi-Uda arrays showing joint frequency-beam reconfigurability, and (c) compact arrays with full steering capabilities. Data obtained from electromagnetic simulations and from \cite{singh2020design}.}
   \label{fig:graphene}
 \end{figure*}

As summarized in Fig. \ref{fig:graphene}, the unique plasmonic properties of graphene in the THz band allows to consider multiple spatial and frequency wireless channels. This is because SPPs are \emph{slow waves} allowing the reduction of the resonant length of the antenna by up to two orders of magnitude \cite{wang2019graphene}. Moreover, SPPs in graphene are tunable electrically, which implies that the resonant frequency of the antenna can be tuned within a wide range by just changing a bias voltage \cite{wang2019graphene, singh2020design}. The tuning dynamics depend on the tuning mechanism, but sub-picosecond speeds have been estimated \cite{wang2019graphene}.
Tunability also implies RF switchability, as antenna elements can be tuned in or out of a certain fixed frequency. These are excellent conditions for the development of ultra-compact, \ac{CMOS}-compatible and digitally programmable transmitarrays able to control both the frequency and direction of radiation without the need for many RF chains \cite{singh2020design}. \hl{Such an approach has been proven through simulation in multiple designs ranging from tunable dipoles, to Yagi-Uda arrays with joint frequency-beam reconfiguration and ultra-dense antenna arrays, as shown in Fig.}~\ref{fig:graphene}. \hlg{In the proposed vision, analog beamforming with single-frontend arrays is envisaged due to evident resource constraints.}

\hl{Graphene has also been exploited to realise RF circuits as amplifiers, power detectors, rectifiers, frequency multipliers, oscillators, mixers, and receivers have been reported}~\cite{Saeed2021}\hl{. The distinct DC characteristics of the graphene-based transistors (GFETs) result in challenges to unleash the full potential of the outstanding electrical properties of graphene in RF circuits. Still, circuits as amplifiers, oscillators, transmitters, \hlg{and even high-frequency phase shifters} are demonstrating performance beyond the expected based on the charge carrier mobility and saturation velocities}~\cite{Saeed2021}. Hence, the employment of graphene-based diodes in RF circuits shows promising performance compared to other well established semiconductor technologies as CMOS and III-V semiconductors~\cite{Saeed2021}.

\subsection*{\centering \scshape \textbf{Wireless Networks within a Computing Package}} %

\begin{figure}
   \centering
   \includegraphics[width=\columnwidth]{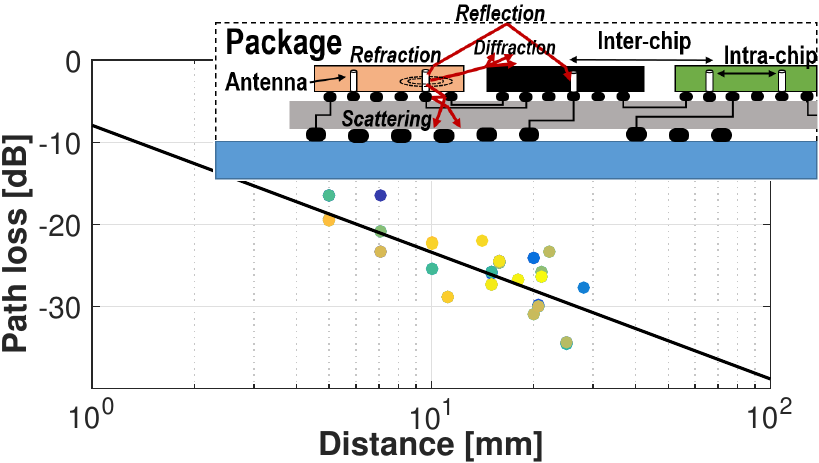}
 	\caption{Path loss as a function of distance in an interposer system composed of four chiplets with four 60-GHz monopoles per chiplet. The inset illustrates the simulated landscape and the different propagation phenomena \cite{Timoneda2018ADAPT}.}
   \label{fig:propagation}
 \end{figure}

The computing package is a new scenario for wireless communications and possesses unique characteristics. Waves propagate \hl{from chip to chip} through a dense environment as illustrated in Fig.~\ref{fig:propagation}, which leads to frequent reflections at the chip interfaces. Moreover, some materials present at the chips such as bulk silicon are lossy. Hence, the channel introduces high attenuation and potentially dense multipath\hlg{, while being in principle not amenable to clean highly directive beam propagation} \cite{abadal2019wave}. Fortunately, this in-package channel is quasi-static and controlled. This implies that the package dimensions and materials can be chosen to reduce both the losses and the \emph{channel length} for a particular set of frequencies \hlg{or phase array distributions}, making it compatible with low-power, ultra-reliable broadband communication \cite{Timoneda2018ADAPT}. Moreover, molecular absorption and \hl{external interference} effects, commonplace in the macro-scale THz networks, are negligible at such short distances and controlled environments \cite{abadal2019wave}.

The use of graphene antennas allows to envision an agile and dense chip-scale wireless network capable of switching between multiple frequencies and transmission modes. Links can be formed within or across chips, using either omnidirectional modes for low-latency broadcast or directive modes with \hlg{controllable and spatially-selective field distributions} for parallel high-speed transmissions. \hl{To implement those modes, the biasing applicable to each graphene antenna element can be obtained offline through accurate electromagnetic characterization and later calibrated with in-package measurements} \cite{singh2020design}. \hl{Then, thanks to the quasi-deterministic nature of the scenario, transmission modes can be expressed as static look-up tables.}

The protocol stack, which in this context would mostly be in charge of determining the most appropriate transmission mode and combination of links, can exploit the \hl{quasi-deterministic} and monolithic nature of the system, whereby the designer has control over the entire architecture. \hl{At the physical layer, the absence of fast-changing channel effects hugely simplifies (if not eliminates) error management and even enables the proactive detection of collisions} \cite{Timoneda2018ADAPT}\hl{. At higher layers,} this allows to eliminate issues such as the \emph{hidden terminal} or the \emph{deafness problem}, where the transmitting and receiving beams are not aligned. Moreover, it provides unique opportunities for \hl{co-design and co-}optimization not available in other scenarios, such as the use of heuristics or pre-established transmission schedules adapted to known applications. Thanks to these opportunities, a simple yet agile protocol stack may withstand the extreme density of the wireless network and the stringent latency deadlines imposed by the system architecture (i.e. less than 10 ns).

\subsection*{\centering \scshape \textbf{Heterogeneous Computing Architectures}} 
\hl{Used as a complement to more traditional wired NoCs/NiPs,} dense wireless networks at the chip scale have the potential to \hl{bypass some longstanding architectural} bottlenecks of computing systems. Historically, these systems have been based on von Neumann architectures where processors and memory are separated and, hence, data communication is key. In this context, the ideal scenario would be having many interconnected cores sharing data through a low-latency, multi-Tb/s, and energy-efficient memory. This, however, is hard to achieve in processors with many cores, which are forced to use complex memory hierarchies to minimize the negative impact of data movement \cite{Nychis2012}. In these processors, the cost of communication increases by an order of magnitude when moving from the nearby small caches to the larger but distant off-chip memories. \hl{Hence, by placing a prohibitive performance burden on functions requiring global communication such as thread synchronization or applications requiring irregular access patterns such as graph algorithms,} these architectures suffer from severe performance, programmability, and scalability limitations.

Disintegration and specialization have been recently adopted to scale performance further, but communication continues being a bottleneck. On the one hand, disintegration implies placing multiple small chiplets on a common substrate and communicating then through high-speed interconnects. However, these interconnects cannot sustain architectures with more than a handful of chiplets due to physical and topological constraints. On the other hand, specialization implies having dedicated engines such as Neural Processing Units (NPUs) for specific workloads like AI. In this case, the custom interconnects are generally not flexible enough to deal with a wide range of applications. Beyond this, disruptive technologies such as in-memory computing are opening the door to chiplets that deliver unprecedented performance by partially alleviating the \emph{memory wall} \cite{Y2020sebastianNatNano}. However, for highly pipelined execution and for being general enough to deal with different applications, an agile communication fabric is essential.

In this context, the vision of agile wireless networks at the chip scale is a very promising approach \hl{towards eliminating the architectural bottlenecks of conventional and novel computing systems}. On classical architectures, wireless networks provide a way to connect cores with distant off-chip memories with similar performance than for on-chip caches. Moreover, their broadcast capabilities open the door to scalable cache coherence, which are necessary to guarantee the programmability of massively parallel processors. On novel architectures based on disintegration and specialization, the flexibility of THz wireless channels allows to improve the connectivity of chiplets while minimizing over-provisioning. These features, coupled with the capabilities of emerging in-memory computing technologies, can be the cornerstone of new breed of general-purpose architectures with accelerator-level performance.

\section*{\Large \textbf{Research Challenges}} 

Albeit promising, the realization of multi-chip architectures integrating wireless nano-networks entails multiple open challenges. We next describe them in a bottom-up approach.

\subsection*{\centering \scshape \textbf{Antenna and Transceiver Implementation}}
\hl{Multiple research groups have simulated graphene antennas in the past years, including resonant, leaky-wave, and even reflectarray antennas} \cite{wang2019graphene,singh2020design}. Simulation and numerical analysis show an attractive reduction of the antenna size, as well as the possibility of tuning the antenna response by only changing the biasing voltage. Unfortunately, until now, no graphene antenna has been manufactured that matches the theoretical predictions. The THz emission of a graphene antenna observed in preliminary experiments is weak, resulting in a very low antenna efficiency. The main reason is that the graphene sheets that constitute the antenna \hl{must have high quality, i.e. ideally a carrier mobility beyond 10\textsuperscript{4} cm\textsuperscript{2}V\textsuperscript{-1}s\textsuperscript{-1},} to achieve resonance. However, while graphene shows such quality \hl{or even higher, in the order of 10\textsuperscript{5} cm\textsuperscript{2}V\textsuperscript{-1}s\textsuperscript{-1}}, as a free-standing layer, it quickly degrades \hl{to a few thousand cm\textsuperscript{2}V\textsuperscript{-1}s\textsuperscript{-1}} when integrated into a THz component or circuit. Multiple approaches have been proposed to remedy this, including the doping of the graphene sheets to improve efficiency, impedance matching between the antenna and the transceiver, or the improvement of the graphene integration process \hl{as described below} \cite{neumaier2019integrating, Saeed2021}.

Besides an antenna, transceiver circuits operating at THz frequencies need to be developed as well \cite{Timoneda2018ADAPT}. Traditional \ac{CMOS} transistors have an increasingly limited performance at this higher frequency range, because it exceeds their \ac{$f_{max}$} and \ac{$f_T$}. To address this issue, heterogeneous technology solutions can be envisioned. Circuits can be implemented in high-frequency technologies like Silicon-Germanium (SiGe), which can operate at 300 GHz, and be co-integrated with the graphene antennas. In order to reach even higher frequencies, \hl{advances in SiGe technology promise to exceed the THz barrier} \cite{Voinigescu2017}. Alternatively, graphene-based active-mixing components could also be established as an alternative for frequency up- and down-conversion with low power consumption \cite{Saeed2021}.

Finally, the implementation of transceivers exploiting the unique properties of GFETs is also a promising yet challenging alternative. Like in the antenna case, the reported \ac{$f_{max}$} and \ac{$f_T$} for GFETs are lower than the expected values due to technology immaturity \cite{Saeed2021}. Consequently, the main challenge comes again from the inability of the GFET to provide the necessary gain at THz frequencies. 
\hl{There are a few solutions for the receiver to overcome this issue, but at the cost of a drastic reduction of the receiver sensitivity} \cite{Saeed2021}. The receiver sensitivity is crucial to compensate the losses of the wireless channel. 
\hl{Another alternative to provide the imperative gain for both transmitters and receivers is to exploit graphene-based devices in a parametric transreactance architecture.} By not relying on the transimpedance of the transistor as in prior works, the device is not limited by the \ac{$f_{max}$} and \ac{$f_T$} and offers the potential to provide the required gain in the THz band.

\subsection*{\centering \scshape \textbf{Technological Integration}} 

Graphene's high carrier mobility, gate-tunable carrier density and general compatibility with silicon technology \cite{neumaier2019integrating} are key to the networking approach laid out in this paper. However, many of its record-breaking performance indicators have been achieved only in highly idealized environments or on champion devices with nearly perfect materials exfoliated manually from natural crystals. Graphene transfer and integration within the chip environment with its dielectrics, electrical contacts and passive components, and at the required performance for wireless in-package communication, remains an open challenge.

In this direction, recent research has achieved the integration of graphene-based RF components such as diodes in a full transceiver circuit \cite{Saeed2021}. However, the target here are the subTHz and THz ranges, which require different, more demanding designs for higher performance. Potentially, quartz-based monolithic integrated circuits would be able to attain the required integration for graphene-based wireless on-chip communication. However, variability-tolerant design still needs to be realized. The main reason is that, in the end, the processes also have to be compatible with future back-end-of-line CMOS integration, especially with respect to process technology and temperatures.

The integration of graphene antennas is even more critical as they not only require significantly higher carrier mobility than transceiver circuits, but also demand higher controllability on its electrical properties to achieve the promised tunability. The doping level of graphene is not yet intentionally controllable, \hl{although various schemes of electrostatic biasing including lateral contacts and graphene-dielectric-metal structures are under continuous improvement} \cite{wang2019graphene}. Even if the doping level can be controlled, there is usually a trade-off between the carrier mobility and the doping level, meaning that high quality and high controllability are hard to achieve simultaneously. In any case, to \hl{attain the required carrier mobility and to} stabilize the performance of the graphene antenna, a delicate encapsulation of the graphene sheet is highly demanded for the graphene antennas. All this requires high-level process and materials engineering and, \hl{while promising approaches such as encapsulation with hexagonal Boron Nitride are being investigated}, integration remains a big challenge for the \hl{relatively} immature graphene-based technology \cite{neumaier2019integrating}.

\subsection*{\centering \scshape \textbf{Models and Protocols for Wireless In-Package Communications}} 

The chip scale is one of the last frontiers in wireless communications and, as such, challenges abound. The scenario exhibits a unique combination of unexplored aspects, stringent constraints (i.e. low area and power) and high performance requirements (i.e. 1--10 ns and 10--100 Gb/s). For instance, the chip-scale THz channel remains largely unexplored, as very few works have studied mmWave and THz wave propagation in enclosed packages \cite{Timoneda2018ADAPT, abadal2019wave}. Clearly, this prevents the direct application of channel models and protocols used in other scenarios and calls for novel, opportunistic, and highly optimized solutions instead. \hlg{For instance, beamforming is clearly affected by the fact that wave propagation occurs in a potentially reverberating channel, which suggests the need for array-package co-design to open a set of spatial channels with affordable antenna apertures.}

In the envisaged scenario, physical layer protocols are challenged by three main constraints. First, coherent schemes are relatively costly due to the need for bulky and power-hungry components such as Phase-Locked Loops (PLLs). Second, the channel is prone to (static) multipath \cite{Timoneda2018ADAPT}. Third, the bit error rate required in this scenario is below 10\textsuperscript{-12}. At upper layers of design, the protocols must manage the reconfigurability properties of the graphene antennas. This represents an open issue as none of the existing wireless chip-scale networks proposals, e.g. \cite{Shamim2017}, support dynamic beam-steering or frequency tuning simply because conventional on-chip antennas cannot have such capabilities. In other words, there are no protocols in the literature that can orchestrate the space-time-frequency channels offered by graphene antennas with the simplicity needed at the chip scale, while adapting to the traffic requirements. An approach where the architecture guides the \ac{MAC} protocol operation, instead of the protocol blindly trying to adapt to the traffic, is promising and unique to this monolithic scenario. 

\subsection*{\centering \scshape \textbf{Massively Parallel Heterogeneous Architectures}} 
The performance, efficiency and flexibility of on-chip and in-package wireless communication is an excellent opportunity to eliminate the bottleneck of current computing systems. The main challenges here are to identify the parts of the architecture where wireless can make a larger impact, to then co-design the network and architecture to optimally exploit the wireless interconnect and caring not to overload it. Often, the challenge resides in the disruptive nature of the potential innovations that wireless can bring in the architecture domain, which has been traditionally driven by rather incremental optimization cycles.

To illustrate this process, let us consider the case of heterogeneous architectures for AI with conventional cores for generic computations and accelerators for certain tasks. In the accelerator side, let us consider an emerging technology such as in-memory computing, which is particularly well suited for deep learning \cite{Y2020sebastianNatNano}. 

In deep learning applications, each layer of a deep neural network can be mapped to an in-memory computing core that stores the corresponding synaptic weights. However, such accelerators present unique challenges when it comes to core-to-core communication and interconnection with conventional digital cores. The physically localized nature of synaptic weights in AI applications entails the need for communication across larger physical distances which could be prohibitive for wired interconnects. Moreover, deep neural networks often feature the need of multicasting data simultaneously from one core to multiple others. In fact, one-to-many connectivity seems to be increasingly prevalent in the most advanced networks, with one clear example being DenseNet, which currently holds the record accuracy for ImageNet classification. In this context, wireless technology brings multiple benefits, but also poses several co-design challenges such as the optimal dimensioning of the wireless network \hlg{and its multiple space-frequency channels}, the development of appropriate multicasting schemes, or the optimal mapping of neurons in the different wirelessly connected in-memory computing cores.

\subsection*{\centering \scshape \textbf{Heterogeneous Simulation Frameworks}} 
The performance evaluation of complex computing systems embedding emerging technologies and architectures is generally a considerable challenge that involves exploring the design space. The use of emerging technologies opens new exploration dimensions and thereby complicates the search, which still needs to be performed in a computationally acceptable time-frame. Moreover, the exploration can only be done if the simulation tools are updated with appropriate performance and cost models of the new technology. 

Relevant to this work, wireless networking is generally poorly supported in computer architecture simulators. Existing tools \cite{Karkar2016, Shamim2017} provide simple estimations of energy, performance, and area trade-offs for simple wireless interconnects in standard architectures. A cross-cutting integration between physical-level and system-level considerations is missing, which prevents the study of thermal effects, radiation and interference patterns, or their impact in the wireless network performance. Therefore, new system simulators are required to guide both application designers in the optimization of their applications, and architects in evaluating complex heterogeneous systems. 

The challenges associated with the development of such a simulator are several. First, the flexibility provided by the frequency-beam reconfigurability and adaptive MAC protocols must be accurately modeled and integrated. Second, system-level aspects such as the impact of transceiver circuits in the thermal profile of the chip, and \emph{vice versa}, must be evaluated and modeled as well. Third, and as mentioned above, wireless communication may lead to disruptive changes in the architecture --and the simulator has to be prepared to cover such cases. This may involve the modeling of emerging technologies and specialized workloads that may be interesting for wireless-enabled architectures, e.g. in-memory computing cores for AI acceleration. In all cases, the simulator must avoid calling complex physical simulators and, instead, rely on accurate behavioral models based on antenna and circuit characterization, wave propagation simulation, protocol design, and architecture development.

\section*{\Large \textbf{Conclusion}} 
Computing system design trends are leading towards multi-chip and accelerator-rich architectures, thus increasing the pressure cast to the already struggling wired interconnects. Wireless interconnects, with their broadcast capabilities and flexibility, have been regarded as a valid complement to the wired backbone, but they fall short in terms of bandwidth. In this position paper, we claim that graphene antennas can make a disruptive impact in future computing systems by virtue of its high bandwidth and beam-frequency reconfigurability. The use of graphene can actually turn a mere wireless interconnect into an agile and dense nanonetwork capable of adapting to the needs of the architecture. For this vision to be realized, however, challenges need to be addressed at multiple levels, from which we highlight the need for (i) optimized methods to integrate the graphene antenna into the chip environment, (ii) establishing methodologies for protocol-architecture co-design, and (iii) integrating wireless communication models into architectural simulators.

\section*{\Large \textbf{Acknowledgment}} 
This publication is part of the Spanish I+D+i project TRAINER-A (ref. PID2020-118011GB-C21), funded by MCIN/AEI/10.13039/501100011033. This work has been also supported by the European Commission under H2020 grants WiPLASH (GA 863337), 2D-EPL (GA 952792), and Graphene Flagship (GA 881603); the FLAG-ERA framework under grant TUGRACO (HA 3022/9-1, LE 2440/3-1), the European Research Council under grants WINC (GA 101042080), COMPUSAPIEN (GA 725657) and PROJESTOR (GA 682675), the German Ministry of Education and Research under grant GIMMIK (03XP0210) and the and the German Research Foundation under grant HIPEDI (WA 4139/1-1).


\begin{IEEEbiographynophoto}{Sergi Abadal} 
is a Distinguished Researcher at Universitat Polit\`{e}cnica de Catalunya (UPC). His research interests include graphene antennas, chip-scale wireless communications, and computer architecture.
\end{IEEEbiographynophoto}


\begin{IEEEbiographynophoto}{Robert Guirado} 
is a graduate student at UPC. His research interests include wireless networks-on-chip, graph neural networks, and antenna design.
\end{IEEEbiographynophoto}


\begin{IEEEbiographynophoto}{Hamidreza Taghvaee} 
is a PhD student at UPC. His main research interests include electromagnetics, metamaterials, and antenna design.
\end{IEEEbiographynophoto}


\begin{IEEEbiographynophoto}{Akshay Jain} 
is a Postdoctoral researcher at UPC, Spain. His research interests include 5G and beyond networks, machine learning, and optimization methods.  
\end{IEEEbiographynophoto}


\begin{IEEEbiographynophoto}{Elana Pereira de Santana} 
is a PhD student at the Institute of High Frequency and Quantum Electronics at University of Siegen. Her research interests include 2-D materials and graphene-based antennas for THz communications.
\end{IEEEbiographynophoto}


\begin{IEEEbiographynophoto}{Peter Haring Bol\'{i}var} 
is Chair for High Frequency and Quantum Electronics, University of Siegen. His research interests include terahertz technology, high frequency electronics, nanotechnology, and photonics.
\end{IEEEbiographynophoto}


\begin{IEEEbiographynophoto}{Mohamed Saeed} 
is a Postdoctoral researcher at the chair of High Frequency Electronics, RWTH Aachen University. His research interests include mixed signal and high frequency circuits in CMOS, III-V, and emerging technologies.
\end{IEEEbiographynophoto}


\begin{IEEEbiographynophoto}{Renato Negra} 
is Professor for High Frequency Electronics at the RWTH Aachen University. His research interests include high frequency systems and circuits in both silicon, III-V, and 2D-material technologies.
\end{IEEEbiographynophoto}


\begin{IEEEbiographynophoto}{Zhenxing Wang} 
is the head of Graphene Electronics Group of AMO GmbH, Aachen. His research interests include electronic, optoelectronic devices and circuits based on graphene and 2-D materials.
\end{IEEEbiographynophoto}


\begin{IEEEbiographynophoto}{Kun-Ta Wang} 
is a PhD student at RWTH Aachen University and research assistant at AMO GmbH, Aachen. His research interests include graphene based antenna and electronic devices.
\end{IEEEbiographynophoto}


\begin{IEEEbiographynophoto}{Max C. Lemme} 
is Professor of Electronic Devices at RWTH Aachen University and Director of AMO GmbH, Aachen. His research interests include electronic, optoelectronic and nanoelectromechanical devices based on graphene, 2-D materials and Perovskites.
\end{IEEEbiographynophoto}


\begin{IEEEbiographynophoto}{Joshua Klein} 
is a doctoral assistant at the École Polytechnique Fédérale de Lausanne (EPFL). His research interests include system-level design, real-time systems, and RISC architectures for machine learning.
\end{IEEEbiographynophoto}


\begin{IEEEbiographynophoto}{Marina Zapater} 
is Associate Professor at HES-SO and external collaborator in EPFL. Her research interests include the design and optimization of heterogeneous architectures, from AI-enabled edge devices to HPC servers.
\end{IEEEbiographynophoto}


\begin{IEEEbiographynophoto}{Alexandre Levisse} 
is a postdoctoral researcher in the Embedded Systems Laboratory (ESL) at EPFL. His research focuses on emerging technologies and architecture design for energy-efficient computing.
\end{IEEEbiographynophoto}


\begin{IEEEbiographynophoto}{David Atienza} 
is Professor of EE and Heads the Embedded Systems Laboratory (ESL) at EPFL. His research focuses on design methodologies for energy-efficient HPC and edge-AI computing for IoT.
\end{IEEEbiographynophoto}


\begin{IEEEbiographynophoto}{Davide Rossi} 
Assistant Professor at the University of Bologna. His research interests include ultra-low power multicore SoC design.
\end{IEEEbiographynophoto}


\begin{IEEEbiographynophoto}{Francesco Conti} 
is Assistant Professor the University of Bologna. His research interests include machine learning applications and accelerators.
\end{IEEEbiographynophoto}


\begin{IEEEbiographynophoto}{Martino Dazzi} 
is a predoctoral researcher at IBM Research Europe. His research interests include machine learning, in-memory computing, and hardware acceleration for machine learning applications.
\end{IEEEbiographynophoto}


\begin{IEEEbiographynophoto}{Geethan Karunaratne} 
is a predoctoral researcher at IBM Research Europe. His research interests include in-memory computing and emerging brain-inspired computing paradigms such as hyperdimensional computing.
\end{IEEEbiographynophoto}


\begin{IEEEbiographynophoto}{Irem Boybat} 
is a postdoctoral researcher at IBM Research Europe. Her research interests include in-memory computing, neuromorphic computing and emerging memory technologies. 
\end{IEEEbiographynophoto}


\begin{IEEEbiographynophoto}{Abu Sebastian} 
is a Distinguished Research Staff Member at IBM Research Europe. His research interests include neuromorphic and in-memory computing, exploratory memory devices and nanoscale science and engineering. 
\end{IEEEbiographynophoto}


\end{document}